\documentclass[5p]{elsarticle}
\usepackage{graphicx,amsmath,amssymb,amstext,amsfonts,color}
\usepackage{caption}
\usepackage{subcaption}

\usepackage{dcolumn}
\usepackage{bm}
\usepackage{color} 
\usepackage[colorlinks,linkcolor=red,citecolor=blue,urlcolor=blue]{hyperref}

\newcommand{\be}{\begin{equation}}

\newcommand{\ee}{\end{equation}}
\newcommand{\ben}{\begin{eqnarray}}
\newcommand{\een}{\end{eqnarray}}
\newcommand{\bes}{\begin{subequations}}
\newcommand{\ees}{\end{subequations}}
\newcommand{\bF}{\begin{figure}}
\newcommand{\eF}{\end{figure}}

\def\tr{ {\rm{Tr }}\,}

\newcommand{\kt}{\rangle}

\newcommand{\ket}[1]{\left|#1\right\rangle}

\newcommand{\mbf}[1]{\mathbf{#1}}
\usepackage{lineno,hyperref}
\modulolinenumbers[5]

\journal{Optics Communications}

\begin{document}

\begin{frontmatter}

\title{Quantum correlations as probes of chaos and ergodicity}


\author{Vaibhav Madhok
\corref{vm}}
\cortext[vm]{Corresponding Author}
\ead{vmadhok@gmail.com}
\author{Shruti Dogra\corref{}}
\author{Arul Lakshminarayan\corref{}}

\address{Department of Physics, Indian Institute of Technology Madras, Chennai, India 600036} 

%

%
%
%
%

\begin{abstract}

Long-time average behavior of quantum correlations in a multi-qubit system, collectively modeled as a kicked top, is addressed here. The behavior of dynamical generation of quantum correlations such as entanglement, discord, concurrence, as previously studied, and Bell correlation function and tangle, as identified in this study, is a function of initially localized coherent states. Their long-time average reproduces coarse-grained classical phase space structures of the kicked top which contrast, often starkly, chaotic and regular regions. Apart from  providing numerical evidence of such correspondence in the semiclassical regime of a large number of qubits, we use data from a recent transmons based experiment to explore this in the deep quantum regime of a 3-qubit kicked top. The degree to which quantum correlations can be regarded as a quantum signature of chaos, and in what ways the various correlation measures are similar or distinct are discussed.

\end{abstract}


\end{frontmatter}


\section{Introduction}

The connections between non-integrability, ergodicity, chaos and entropy production form the cornerstone of classical statistical mechanics. A key goal of quantum information theory is to extend this to the quantum domain and explore these connections in the light of information theory. For example, the dynamical generation of entanglement and other non-classical correlations  \cite{Miller/Sarkar,Lakshminarayan,Lakshminarayan/Bandyopadhyay2002, Ghose, Wang2004, tmd08,  mgtg15}, thermalization in closed quantum systems \cite{Deutsch91, Srednicki94} and its connections to non-integrability and chaos have been extensively studied theoretically and experimentally, especially for finite dimensional systems like the kicked top \cite{Chaudhary,Neill16}. Most of these studies have been undertaken in the semi-classical limit where high entanglement and quantum discord production
is attributed to generation of pseudo-random states and random matrix theory \cite{Lakshminarayan,Lakshminarayan/Bandyopadhyay2002,  tmd08, mgtg15}. As classically chaotic dynamics takes initially localised distribution and spreads them over phase space, the corresponding quantum dynamics takes initially localized coherent states to pseudo random states in the Hilbert space.
However, for systems far below the semiclassical limit, it has been recently argued that the correspondence between entanglement and classical phase space features
occurs through symmetries of the system and how much the states gets spread out over the phase space \cite{Arjendu17}. 

In this work, we provide numerical as well as experimental evidence for various quantum signatures of classical chaos and report a new correspondence between long time average of Bell correlation function, as defined below, and
structures in the classical phase space. Contour plots of long time average entanglement, quantum discord, tangle, concurrence and the Bell correlation functions, all show features of the classical phase space.
 We also provide evidence of this correspondence in the deeply quantum regime of a 3-qubit kicked top that is based on data from a recent experimental realization with superconducting qubits \cite{Neill16}. It is remarkable and surprising that the 3-qubit kicked top is exactly solvable \cite{Lakshminarayanprivate, SDograprep}, this raises very interesting questions on the connections between ergodicity, integrability and chaos in the quantum domain. We conclude our work with a discussion of these directions.

\section{Background}
The quantum kicked top Hamiltonian \cite{Haake, Chaudhary} is given by
\begin{equation}
\label{Eq:QKT}
H=\frac{\kappa}{2j}{J_z}^2 \sum_{n = -\infty}^{ \infty} \delta(t-n\tau)+\frac{p}{\tau} \, {J_y}.
\end{equation}
Here $J_x, J_y$ and $J_z$ are components of the angular momentum operator $\mbf{J}$. The time between periodic kicks is given by $\tau$. The Floquet map is the unitary operator
\begin{equation}
\label{uni}
U = \exp\left ({-i \frac{\kappa}{2j \hbar} J_z^2} \right)\exp\left({-i \frac{p}{\hbar} J_y}\right), 
\end{equation}
which evolves states just after a kick to just after the next, that is over
one time period $\tau$. 
The parameter $p$ measures rotation about the $y$ axis, and  $\kappa$ is the magnitude of a twist applied between kicks and controls the degree of chaos in the system. In the following we set $\hbar=1$ and $p=\pi/2$.

If the initial state is $|\psi(0)\kt$, the state after $n$ kicks is given by $U^n |\psi(0)\kt$. In order to study the correspondence of the quantum dynamics with the classical phase space, we take the minimum uncertainty wave packets as the initial
states of quantum evolution.
Such states are the spin coherent states, which can be expressed as \cite{Glauber, Puri}
\begin{equation}
\label {SCS}
\ket{\theta,\phi} = R(\theta,\phi) |j,j\rangle; -\pi \leq \phi < \pi, 0 \leq \theta \leq \pi
\end{equation}
where,
\begin{equation}
\label {R}
 R(\theta,\phi) = \exp \{ i\theta [J_x \sin \phi - J_y \cos \phi]\} 
\end{equation}
with the expectation value of $\bf J$ given by
\begin{equation}
\label {R}
 \langle \theta,\phi | \mbf{J}/j | \theta, \phi \rangle  = (\sin \theta \cos \phi, \sin \theta \sin \phi, \cos \theta).
\end{equation}

The single large spin-$j$ top can be considered as a collective spin of $2j$ spin-$1/2$ particles or qubits.
Using  $J_i =\sum_{k=1}^{2j} \sigma^i_{k}/2$ where $i=x,y,z$ the unitary operator in Eq.~(\ref{uni})
can be considered as that of $2j$ qubits. Thus a single large top is mapped to a model of many qubits, and both
theoretical work and experimental realizations have exploited this feature. However a very important consequence
of the mapping is that one is restricted to a permutation symmetric subspace ($2j+1$ dimensional) of the $2^{2j}$ dimensional 
multi-qubit space. The correlation measures that are explored are those of the multi-qubit system. 

Each of these qubits are initialized in the same state 
($|\psi_0\kt=\cos \frac{\theta}{2} |0\kt + e^{i \phi} \sin \frac{\theta}{2} |1\kt$, 
in the computational bases), such that the initial state of the
$2j-$ qubit system is the spin coherent state in Eq.~(\ref{SCS}), $ |\psi(0)\kt = |\theta, \phi \kt =\otimes^{2j} |\psi_0\kt$.
The time evolution restricts the evolved states at later times to also be permutation symmetric, but these are
naturally states with non-trivial correlations. 



\section{Measures of quantum correlations}

For completeness we recall the definitions of various metrics of quantum correlations that we use subsequently.

\subsection{Entanglement entropy and concurrence}

The entanglement measure usually considered is the von Neumann entropy associated with the reduced density matrix
of a single qubit, $S = - \text{Tr} \rho_{k} \log (\rho_{k})$,
where $\rho_{k}$ is the density matrix of the the qubit $k$, {\it i.e.}, we are interested in the entanglement between this qubit and the rest.
In this paper we use the linear entropy instead of von Neumann, and this define as $S_l=1-\text{Tr}(\rho_k^2)$, but this is qualitatively similar. As the complete state is permutation symmetric this is unique and independent of which qubit is considered, hence there is no need to index the entropy. 

The entanglement between qubits $k$ and $l$ is measured by its concurrence $\mathcal{C}$ that is simply calculated from the two qubit reduced
density matrix $\rho_{kl}$:
\begin{equation}
\mathcal{C}=\max(0, \sqrt{\lambda_1}-\sqrt{\lambda_2}-\sqrt{\lambda_3}-\sqrt{\lambda_4})
\end{equation} 
where $\lambda_i$ are the eigenvalues in decreasing order of the matrix $\rho_{kl} (\sigma_y 
\otimes \sigma_y)\rho_{kl}^{\ast}(\sigma_y \otimes \sigma_y)$. Here $\ast$ denotes complex conjugation in the standard $\sigma_z$ basis,
and once again there is no need to label the concurrence as due to permutation symmetry every pair of qubits are identically entangled.

\subsection{Quantum Discord}

Quantum discord aims to capture all the quantum correlations in a bipartite quantum system~\cite{oz02,oz02-2}. To accomplish this, we remove the classical correlations from the total correlations in a system. Consider,  $\rho_{AB}$, then the quantum mutual information is $\mathcal{I}(A:B) = \mathcal{H}(A) + \mathcal{H}(B) - \mathcal{H}(A,B),$
where $\mathcal{H}(X)$ is the von Neumann entropy of $\rho_X$.
However, the mutual information for classical probability distributions has another definition given by the Bayes' rule, namely $I(A:B) = H(A)-H(A|B).$
Here, the conditional entropy $H(A|B)$ is the average of the Shannon entropies of $A,$ conditioned on the values of $B$, and quantifies the ignorance in $A$ given the state of $B$. 

In the quantum case, we can describe measurements on $B$ by a POVM (positive-operator valued measure)  set $\{\Pi_i\}$. Maximizing the consequent quantum mutual information over all $\{\Pi_i\},$ we obtain 
\begin{equation}
\mathcal{J}(A:B) = \max_{\{\Pi_i\}}(\mathcal{H}(A) - \tilde{\mathcal{H}}_{\{\Pi_i\}}(A|B)).
\end{equation}
Hence we arrive at a definition for quantum discord as the difference of two ways of defining the quantum mutual information $\cal{I}(A:B)-\mathcal{J}(A:B)=$
\begin{equation}
\label{discexp}
\mathcal{D}(A:B) = \mathcal{H}(B)-\mathcal{H}(A,B)+ \min_{\{\Pi_i\}}\tilde{\mathcal{H}}_{\{\Pi_i\}}(A|B),
\end{equation}
with $\{\Pi_i\}$ being rank-1 POVMs.  Quantum discord is non-negative for all quantum states~\cite{oz02,oz02-2,dattathesis}, and it is subadditive~\cite{md10}.

\subsection{3-Tangle}
The 3-tangle is a measure of genuine tripartite entanglement defined for a pure state of three qubits \cite{wootterstangle}.
For a three qubit pure state $\rho_{ABC}$ (with its two-party reduced states 
$\rho_{mn}$ and single-party reduced states $\rho_{m}$, where $m,n \in \{A, B, C \}$), the measure of 3-tangle is given by
\be
\tau({\rho_{A}:\rho_{B}:\rho_{C}}) =\tau({\rho_{A}:\rho_{BC}}) - \tau({\rho_{AB}}) - \tau({\rho_{AC}}).
\ee
First term on the right hand side quantifies the entanglement of $\rho_A$ with rest of the system ($\rho_{BC}$), 
and is given as $\tau({\rho_{A}:\rho_{BC}})=2(1-\text{tr}(\rho_A^2))$, while second and third terms 
are the squares of the concurrences between the the parties in the parenthesis.

\subsection{Bell Correlation function}
Quantum mechanics permits correlations that are much stronger than classical mechanics, i.e., that cannot be accounted by any local hidden variable theory \cite{nielsen-book-02}. Such nonlocal correlations can be quantified in ways that are somewhat different from entanglement and thus is of interest. It maybe pointed out that this is not studied yet in contexts involving classical chaos.

Defining $Q=\sigma_{z} $, $R = \sigma_x$, $S= (-\sigma_x - \sigma_z)/\sqrt{2} $ and  $T= (\sigma_x - \sigma_z)/\sqrt{2}$ where $\sigma_i$ are the usual Pauli operators, the (absolute value of) expectation value $| \langle QS  +RS+RT-QT  \rangle|$ with respect to a given two-qubit state is a measure of non-locality. However unlike other correlation measures above, this is {\it not} invariant under local transformations, as these change the measurement settings. One may seek to find optimal settings for the largest possible correlation and this was done in \cite{Horodecki-pla-1995, Bartkiewicz-2013}. The setting is now state dependent and the maximum correlation is related to ${\cal M}(\rho)=\text{max}_{i<j} h_i+h_j$, where $h_i$ are eigenvalues of 
the $3 \times 3$ correlation matrix $T T^{\dagger}$ and $T_{ij} =\tr(\sigma_i \otimes \sigma_j \rho)$. If ${\cal M}(\rho)>1$ (we will refer to this as ``Bell correlation function", and this is invariant under local unitaries) there is some measurement setting that violates the CHSH inequality. It is easy to see that ${\cal M}(\rho)=0$ for the most mixed state, while it is $=1$ for all pure unentangled states and $=2$ for maximally entangled ones.

\section{Dynamics of Quantum Correlations}

\subsection{Quantum signatures of classical chaos: Semiclassical limit}

Quantum signatures of classical chaos occur in various contexts. Level statistics and connections to random matrix theory \cite{Haake},
quantum information theoretic signatures \cite{Lakshminarayan/Bandyopadhyay2002, Lakshminarayan, tmd08, Ghose, Miller/Sarkar, Wang2004, mgtg15, mrgi14}, hypersensitivity of dynamics under perturbations \cite{per00, sc96}, 
and those related to open system dynamics and the rate of decoherence \cite{Zurek/Paz}.
In the signatures based on quantum correlations, entanglement, discord and to some extent, concurrence \cite{LakSub2003, ScottCaves2003}, have been studied. 

\begin{figure}[h]
\includegraphics[scale=1.0]{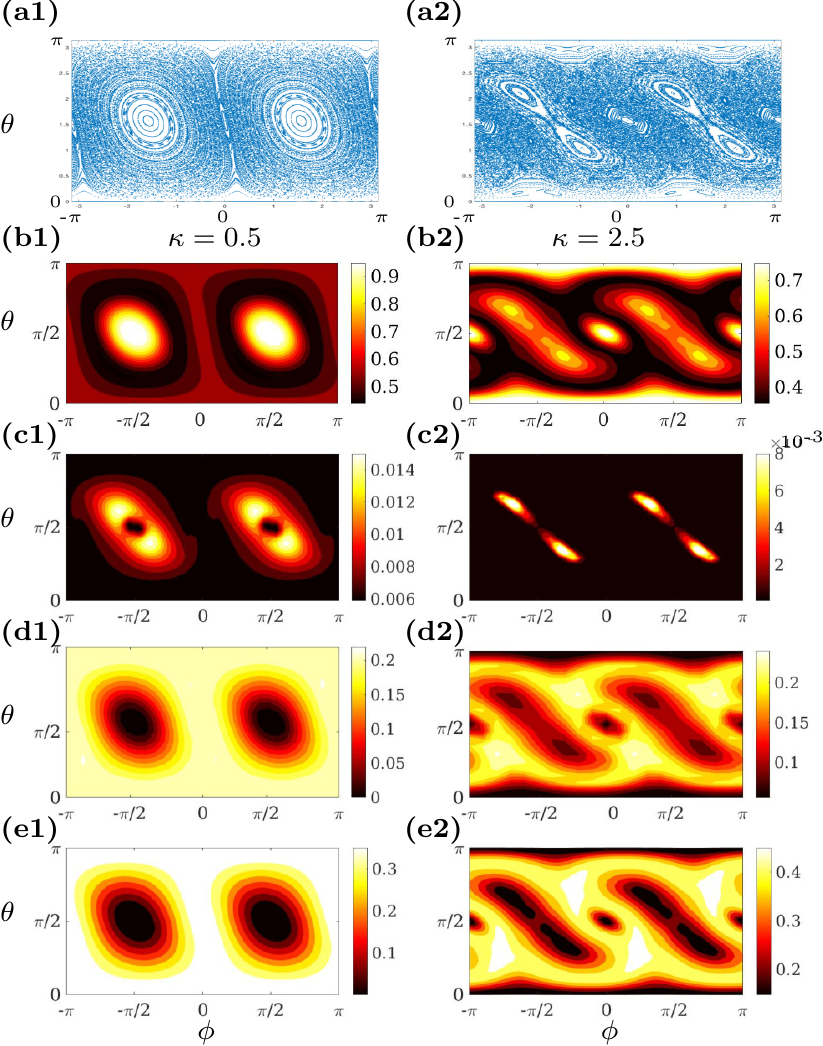}
\caption{Columnwise comparison showing dynamically generated quantum correlations as superb signature of classical chaos in a regular ( $\kappa=0.5$) and mixed  phase spaces ( $\kappa=2.5$), (a1 \& a2) Classical phase space, Poincar\'{e} sections, (b1 \& b2) Long time average of Bell correlation function, (c1 \& c2)  Long time average of Concurrence, (d1 \& d2) Long-time average discord, (e1 \& e2) Long-time average linear entropy - all as a function of mean coordinate of the initial projected coherent state. Spin $j = 20$ and average is taken over 100 kicks.}

\label{F1}

\end{figure}

Here we give numerical evidence of other information theoretic quantities, like the Bell correlation function and concurrence and tangle whose behavior exhibits quantum signatures of classical chaos. The signatures of chaos in the evolution of these quantities is evident in the long time-averaged values as we scan through different initial conditions. Figure~(\ref{F1}) displays remarkable similarities between structures in the classical mixed phase space and the time average of  quantum correlations for a relatively large value of spin, $j = 20$, which corresponds to 40 qubits.

Chaotic initial conditions generally go to a higher average value than regular initial conditions for entanglement and discord, while concurrence and Bell function ${\cal M}(\rho)$ have an opposite tendency. The values for all initial conditions in the chaotic sea saturate to nearly the same average behavior. The concurrence is smaller for initial states launched from chaotic parts of the phase space than from regular. This is understood to be be due to the development of multipartite entanglement, which implies that correlations between any two qubits decrease. This is also a reflection of the unique monogamy of quantum correlations that forbid sharing of large correlations between multiple pairs of particles. In part c1 of Fig.~(\ref{F1}) we also notice that the concurrence is very small at the center of the regular island. Of course in this case there is no chaos that is lowering this value, but rather the lack of any entanglement at all, due to the stable fixed point nature of the classical dynamics. 

The function ${\cal M}(\rho)$ shows a very similar pattern as the concurrence, except that the evolution at the center of the regular islands for $\kappa=0.5$ does comparable values as other regular orbits, in contrast to the concurrence. This can be understood as a smaller value of ${\cal M}(\rho)$ occurs for mixed separable states 
as generated in a chaotic evolution, than for a state that is nearly pure,
as the states localized at fixed points must be.
 
All this can be attributed to generation of pseudo random states, in the permutation symmetric subspace, due to the chaotic dynamics.
 Quantum chaotic dynamics leads to a generation of states that are a superposition of basis states in a tensor product vector space of the parties involved and resembles those sampled randomly according to the appropriate Haar measure \cite{Haake}.  In the limit of high dimensional Hilbert space, the semiclassical limit, the saturation values attained for quantum correlations due to chaotic dynamics can be well predicted by random matrix theory \cite{ tmd08, Lakshminarayan/Bandyopadhyay2002, Lakshminarayan, wootters2}. These predictions can be extended to mixed phase spaces with regular islands immersed in a chaotic sea.  Trail et al., invoked  Percival's conjecture \cite{Percival, tmd08} that separates eigenstates into chaotic and regular classes and explained the generation of entanglement in mixed phase space to be that of a random state in a chaotic subspace of the total Hilbert space. However two small sub-systems of pseudo-random pure states have little entanglement or non-locality associated  them and is consistent with the behavior of concurrence and the Bell correlation function.
 
%
%

\subsection{Quantum correlations and ergodicity in deeply quantum regime}

\begin{figure}[h]
\includegraphics[scale=1.0]{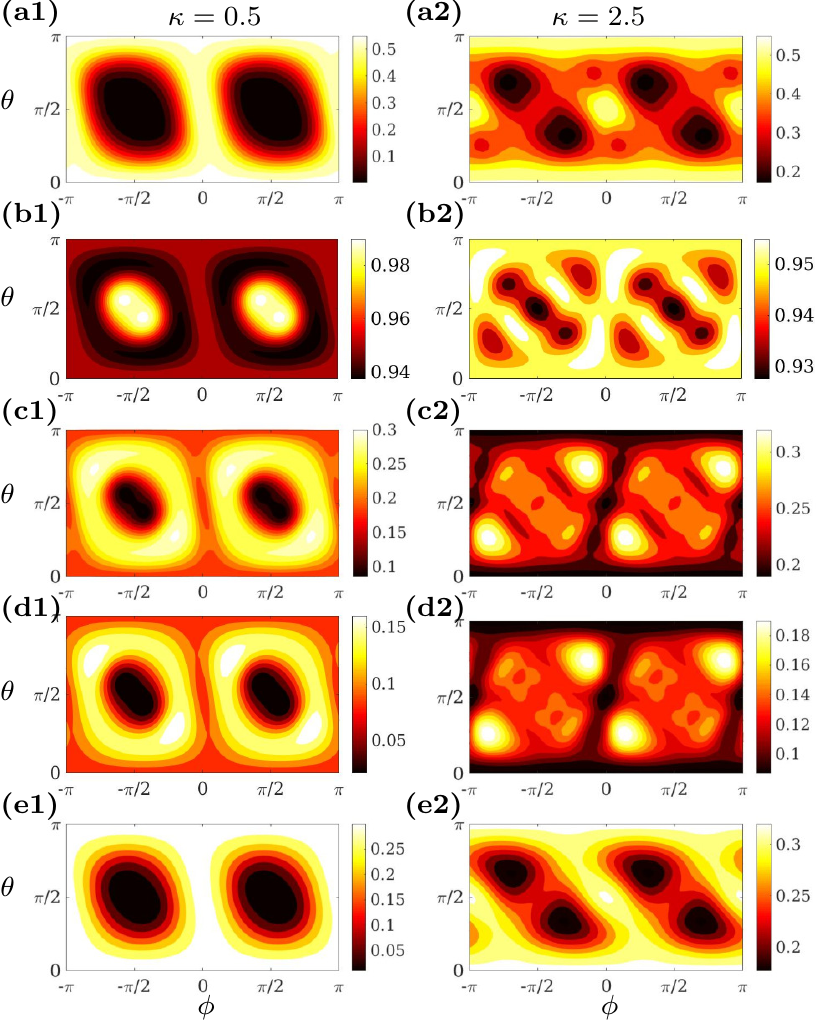}
\caption{Columnwise comparison showing dynamically generated quantum correlations as superb signature of classical chaos in a regular ( $\kappa=0.5$) and mixed  phase spaces ( $\kappa=2.5$), (a1 \& a2) Long time average tangle, (b1 \& b2) Long time average of Bell correlation function, (c1 \& c2)  Long time average of Concurrence, (d1 \& d2) Long-time average discord, (e1 \& e2) Long-time average linear entropy - all as a function of mean coordinate of the initial projected coherent state. Spin $j = 3/2$ and average is taken over 100 kicks.}
\label{F2}

\end{figure}

In order to study the quantum classical correspondence between classical phase space and quantum correlations deep in the quantum regime, we take the case of $j = 3/2$, the case of the 3-qubit kicked top. This is especially motivated as the experimental data we will analyze is for the case of 3 qubits. Remarkably, we find in Fig.~(\ref{F2}) that quantum dynamics even in this extreme quantum limit as features that resemble the classical phase space structures.
In fact even for the case of just $j=1$, or 2 qubits, this has been observed and discussed in \cite{Arjendu17}. In the previous literature, this has been attributed to the ergodic nature of  dynamics \cite{ Arjendu17, Neill16, matzkin}. 

However there are some notable differences between this and the large $j$ cases. There is an extra plot showing the 3-tangle average as well, which is a well-defined measure of tripartite entanglement for three qubit pure states as we have in this case.  It is seen that there is a striking resemblance between the tangle and the entanglement, and both still resemble to a large extent a coarse grained classical phase space. On the other hand now discord joins both concurrence and the Bell correlation function and shows peculiar structures for the case of $\kappa=2.5$. In this deep quantum regime there is significant amount of entanglement between two qubits and entanglement has not entirely transformed into multipartite measures. 
A more detailed understanding of these structures is lacking at the moment.

We analysed the experimental data (provided generously by the authors of ~\cite{Neill16}) based on a 
system of three superconducting qubits to study the dynamics of 
these quantum correlations. Choosing different values of $\theta$ and $\phi$, the three-qubit system is experimentally initialised in various different coherent states, and allowed to evolve for 20 time steps (as per Eq.(\ref{uni})). The final state at the end of each time step is
experimentally reconstructed via quantum state tomography (details are given in a future publication). These states are further used to calculate various quantum correlations as a function of time.

The experiment also considers the two values of the chaoticity parameter, $\kappa = 0.5$ and $\kappa = 2.5$  as we apply the Floquet operator on the \textit{same} initial state.
  We see a change in the dynamics of quantum correlations as the parameter, $\kappa$, takes the classical dynamics from a regular to a mixed phase space. As we increase the degree of chaos in the system, the initial state under consideration transitions from being in the regular island, to be on the border between the chaotic sea and regular regions. 

\begin{figure}[h]
 \centering
 \includegraphics[scale=1.0,keepaspectratio=true]{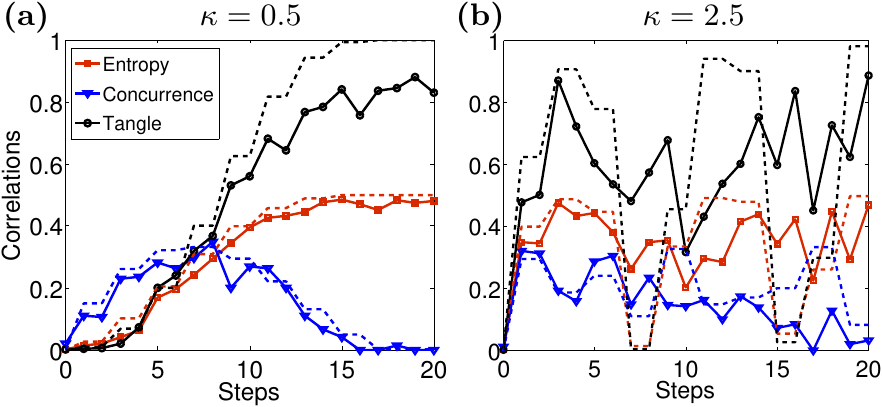}
 \caption{This figure shows the dynamics of linear entropy, 
 concurrence and 3-tangle of a system of three qubits initialised 
 in state $|000\rangle$ under a kicked top Hamiltonian.
 Plots corresponding to the chaoticity parameter, (a) $\kappa=0.5$ and 
 (b) $\kappa=2.5$ are shown. Solid red curve with squares correspond to the experimentally
 obtained values of linear entropy of a single-party reduced density matrix, 
 solid blue curve with triangles stands for the experimental concurrence 
 values, and solid black curve with circles is the experimental 3-tangle.
 Dashed curves in red, blue and black colors are the simulated values 
 of linear entropy, concurrence and 3-tangle respectively.}
\label{F3}

\end{figure}
\begin{figure}[h]
 \centering
 \includegraphics[scale=1.0,keepaspectratio=true]{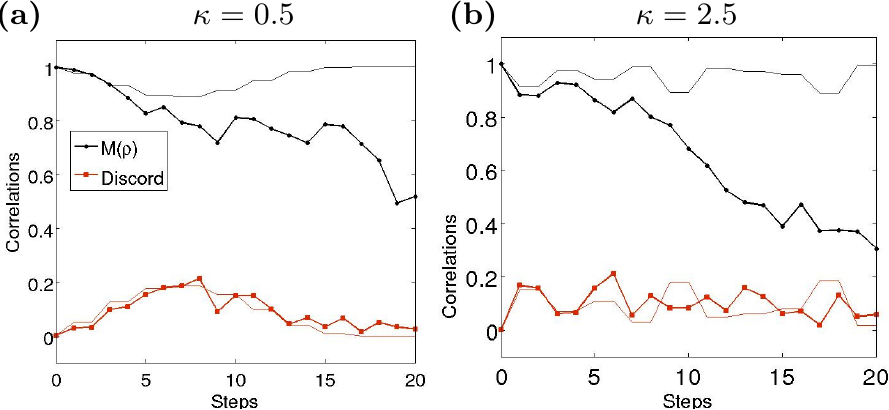}
 \caption{This figure shows the dynamics of Discord and {Bell correlation function ($M(\rho)$)} under the kicked top Hamiltonian with different strengths, (a) $\kappa=0.5$ and (b) $\kappa=2.5$. Thick solid red curve with squares correspond to the experimentally obtained values of Discord and thick black curve with circles stands for the experimental $M(\rho)$ values. Thin curves in red and black colors are the simulated values of discord and $M(\rho)$ respectively. Three qubit system was initialised in the state $|000\rangle$.}
\label{F4}   
\end{figure}
Figures \ref{F3} and \ref{F4} show the behavior of entanglement (as captured by the linear entropy), discord, concurrence, tangle and the Bell correlation function for the case of regular and mixed phase spaces. For the case of regular phase space, the rate of growth of all quantum correlations is gradual whereas the initial rate of growth is rapid in case of the mixed and chaotic phase space.The rapid growth indicates signatures of classical Lyapunov exponents. Therefore, for a fixed family of maps, the generation of correlations is influenced by the degree of chaos in the system. The simulations in Figure  \ref{F3} show an alternating ``staircase" like behavior for correlations which has a remarkable confirmation in experiment as shown in the same figure. This is becasue the dyamics at such points essentially becomes local and does not cause a change in the value of  correlations. We have a complete analytical explanation for the above and this will be reported elsewhere \cite{Lakshminarayanprivate, SDograprep}. The behavior of correlations also reflect interesting dynamics. Initially, we start with a product state with no correlations. Linear entropy, concurrence and tangle all start to rise. Linear entropy captures the correlation between a single qubit and the ``bath" and continues to rise as the single qubit state becomes increasingly mixed. This is in contrast to concurrence that rises initially but as the two qubit state starts becoming increasingly mixed, the concurrence starts to decrease and drops to zero. 
The value of tangle captures how the correlations get spread out in multiple parties, i.e, from bipartite correaltions as captured by concurrence to multipartitie correlations
given by the tangle.
Figure \ref{F4} shows the analogous behavior of the Bell correlation function for the same initial state with varying degrees of chaos. It is to be noted that the value of ${\cal M}(\rho)$ never exceeds $1$ and therefore no measurement setting will violate the CHSH inequality in this scenario. This is a consequence of the monogamy of ${\cal M}(\rho)$ and the fact that the state is restricted to the permutation symmetric subspace at all times. 
 Since at most one of $\rho_{12}$ and  $\rho_{13}$ can violate the CSHS inequality and the value of ${\cal M}(\rho)$ is same regardless of which two qubits of a three qubit permutation symmetric state are taken to construct $\rho$, neither of them violates it \cite{Toner59, ramanathan-prl-2011, sharma-pra-2016}.
The connection of Bell correlation values to non-locality and chaos is intriguing and will be subject of future studies.

\section{Discussion}

On the face of it, there seems to be an obvious connection between purely quantum aspects of multipartite systems like entanglement, concurrence, discord and Bell correlation functions and chaos. Chaos arises due to coupling of degrees of freedom. In the quantum domain, coupling of degrees of freedom can potentially lead 
to generation of correlations. The above reasoning, though compelling, is not the whole story. The different behaviour of dynamics of correlations in the regular regions and chaotic sea of a mixed phase space is intimately related to the generation of pseudo-random states in the joint Hilbert space of the system. 

It has been recently found that, unlike previously thought \cite{Arjendu17}, the 3 - qubit kicked top is solvable analytically. We have \cite{Lakshminarayanprivate, SDograprep} recently found the analytical solution to this problem. This opens a range of interesting questions regarding the connections of thermalization and integrability \cite{Neill16} and chaos in such systems. To what extent closed quantum systems, that are integrable in the sense that an analytical solution exists quantum mechanically but whose classical limit is chaotic, thermalize? Another intriguing avenue is the issue of quantum-to-classical transition in such systems and the role of quantum correlations here. 
Under what conditions does a quantum trajectory that tracks a measurement of a given observable follow the classical trajectory? 
Habib et al. \cite{bhj03} have shown that one can recover classical trajectories from quantum systems when the quantum system is continuously measured with appropriate measurement strength. In particular, if the measurement strength is strong enough to localize the wave packet such that the interference across different ``paths" is negligible, but weak enough so that the back-action noise is small, the system follows classical trajectories.
 It has also been shown that entanglement between the two sub-systems causes strong measurement back-action and one does not recover classical trajectories in this regime. It is important for both subsystems to be sufficiently macroscopic for the recovery of classical trajectories. 
 Furthermore, the Ehrenfest ``break time", the time at which quantum expectation values and classical motion start to deviate from each other,  
 is exponentially small for chaotic systems as compared to regular systems. Therefore, quantum chaotic multipartite systems are a very interesting avenue to study interplay of chaos, quantum correlations and measurements and their role
in quantum-to-classical transition. The systems, like the kicked top that we consider are just a start.
Moreover, experimental control over such systems in the superconducting qubits set up, like the one we have considered, is a huge challenge but becoming increasingly feasible. Perhaps even more interesting is the connections between chaos, integrability and information as captured by correlations.  As the initial state, a coherent state on the entire multi-qubit Hilbert space of the system, is varied from the regular to the chaotic regions of the corresponding classical phase space, a contour plot of the long time averaged correlations  reproduces the structures of the classical stroboscopic map. One import from this is that as classical chaos produces classical information, captured by uncertainty and Lyapunov exponents, quantum chaos
 generates quantum information that is manifest is superposition of quantum states of a complex quantum
system. Quantum correlations in these pseudo random states are one way to characterize and quantify this information and superpositions and hence
are signatures of the underlying chaotic dynamics. That these signatures persist in a deeply quantum regime of 3 qubits in a system that is integrable in the quantum sense
needs further investigation.

\section{Acknowledgement} We thank the research group of John Martinis for providing the experimental data from \cite{Neill16}.



 \section*{References}
\bibliographystyle{elsarticle-num}
\bibliography{discord}

\end{document}